\documentclass[9.5pt,journal,compsoc]{IEEEtran}

\ifCLASSOPTIONcompsoc
  \usepackage[nocompress]{cite}
\else
  \usepackage{cite}
\fi

\usepackage{graphicx,lipsum,afterpage,comment}
\usepackage{cite}
\usepackage{amsmath,amssymb,amsfonts}
\usepackage{algorithmic}
\usepackage{graphicx}
\usepackage{textcomp}
\usepackage{xcolor}
\usepackage{mathtools}
\usepackage{xspace}
\usepackage{comment}
\usepackage{float}
\usepackage{caption2}
\usepackage{graphicx,url}
\usepackage{wrapfig}

\ifCLASSINFOpdf
\else
\fi

\usepackage{amsmath}

\usepackage[english,ruled,linesnumbered,vlined]{algorithm2e}
\usepackage{enumitem}
\usepackage{multirow}
\usepackage{bigstrut}
\usepackage{mathtools}
\usepackage{boldline} 
\usepackage{comment}
\usepackage{xcolor}
\usepackage{textgreek}
\usepackage{silence}
\usepackage{graphicx}
\usepackage{booktabs}

\usepackage[utf8]{inputenc}

\usepackage{color}

\usepackage{tikz}

\newcommand\copyrighttext{%
  \footnotesize This work has been submitted to the IEEE for possible publication. Copyright may be transferred without notice, after which this version may no longer be accessible.}
\newcommand\copyrightnotice{%
\begin{tikzpicture}[remember picture,overlay]
\node[anchor=south,yshift=10pt] at (current page.south) {\fbox{\parbox{\dimexpr\textwidth-\fboxsep-\fboxrule\relax}{\copyrighttext}}};
\end{tikzpicture}%
}

\newcommand\Func[2]{%
    function(#1)\{%
    \algocf@group{#2}%
    \}\;%
}

\usepackage{array}

\ifCLASSOPTIONcompsoc
  \usepackage[caption=false,font=footnotesize,labelfont=sf,textfont=sf]{subfig}
\else
  \usepackage[caption=false,font=footnotesize]{subfig}
\fi

\usepackage{stfloats}

\usepackage{url}

\newcommand{\hsep}{\hspace{3 mm}}

\begin{document}

\title{E-Health Sensitive Data Dissemination Exploiting Trust and Mobility of Users}

\author{{\bf Agnaldo de Souza Batista}, {\bf Michele Nogueira}, {\bf Aldri Santos}
\thanks{A.S. Batista, M. Nogueira and A. Santos are with the Federal University of Paran\'a, Brazil. Email addresses: \{asbatista,michel.nogueira,aldri\}@ufpr.br
\protect\\ Manuscript received []; revised [].}}

\markboth{IEEE TRANSACTIONS ON MOBILE COMPUTING,~Vol.~[], No.~[], Month~Year}%
{Batista \MakeLowercase{\textit{et al.}}: E-Health Sensitive Data Dissemination Exploiting Trust and Mobility of Users}

\IEEEtitleabstractindextext{
\begin{abstract}

E-health services handle a massive amount of sensitive data, requiring reliability and privacy. The advent of new technologies drives e-health services into their continuous provision outside traditional care institutions. This creates uncertain and unreliable conditions, resulting in the challenge of controlling sensitive user data dissemination. Then, there is a gap in sensitive data dissemination under situations requiring fast response (e.g., cardiac arrest). This obligates networks to provide reliable sensitive data dissemination under user mobility, dynamic network topology, and occasional interactions between the devices. In this article, we propose STEALTH, a system that employs social trust and communities of interest to address these challenges. STEALTH follows two steps: clustering and dissemination. In the first, STEALTH groups devices based on the interests of their users, forming communities of interest. A healthcare urgency launches the second, in which STEALTH disseminates user sensitive data to devices belonging to specific communities, subjected to the level of trust between devices. Simulation results demonstrate that STEALTH ensures data dissemination to people who can contribute toward an efficient service. STEALTH has achieved up to 97.14\% of reliability in accessing sensitive data with a maximum latency of 170 ms, and up to 100\% of availability during emergencies.

\end{abstract}

}

\maketitle
\copyrightnotice

\IEEEdisplaynontitleabstractindextext

\IEEEpeerreviewmaketitle

\IEEEraisesectionheading{\section{Introduction}\label{sec:introduction}}

\IEEEPARstart{T}{he} Internet allows us to access an increasing number of online services, supporting the population in different application domains like healthcare, transportation, surveillance, among other. Forecasts estimate that the size of digital health market can exceed USD 504.4 billion by 2025~\cite{health2019market}. In this context, \mbox{e-health} services, like online patient monitoring, continuous glucose monitoring, automated insulin delivery, and electronic health records, have leveraged citizens’ quality of life streamlining care and contributing to reduce operational costs. These services collect and disseminate data often through opportunistic contacts between geographically near devices, when interactions enable the communication~\cite{garyfalos2008coupons}. Data dissemination relies on sharing, being a demanding task, intensified by a high data dissemination frequency, location, and content~\cite{sudhindra2014,umarani2013survey}. Many services require dynamically established local or global networks to support their operation under mobility. 

Mobile devices collect various types of data~\cite{ruiz2014analysis}, allowing improvements in different domains. The interaction between people and smartphones has intensified, allowing the formation of temporary local networks, where data are exchanged for different purposes and usually for a period. In general, these dynamics rely on structured wireless networks (e.g., structured WiFi-based or cellular networks), which offer extensive coverage inside buildings and cities. These networks inhibit direct communication between devices, which often impacts the response time on critical events (e.g., traffic accident, health emergency, and environmental disaster). \mbox{E-health} services lean on well-structured networks in hospitals and clinics to disseminate sensitive data. However, people also might need these services anywhere and anytime, including exterior environments.

People can be suddenly stricken by a health issue.  In 2011, more than \mbox{326 200} people experienced medical emergencies observed  out-of-hospital in the United States~\cite{mozaffarian2015executive}. Therefore, building and maintaining urban outside environments (e.g., streets and avenues) are a challenge, mainly to provide \mbox{e-health} services, given the need for a reliable network infrastructure. The nature  of sensitive health data addressed by \mbox{e-health} services (e.g., vital data, blood pressure information, glucose measurements, exam results, and medical prescriptions) requires reliability in data dissemination and protection from unauthorized access. Furthermore, due to their emergency nature, medical alerts must be promptly transmitted~\cite{movassaghi2014wireless} with a maximum latency of 125ms~\cite{ieee2012}, once the consequences in the face of losses and delays can be severe to users health~\cite{latre2011survey}. 

Currently, the existing mechanisms are inadequate to deal with healthcare urgency in urban environments, such as a cardiac arrest, change in blood pressure, or change in glucose level, because they require prompt treatment. Safety services deliver data to the right people and prevent leaks. Several works in the literature address data availability in unstructured networks in the contexts of Internet of Things (IoT)~\cite{al2017trust, bao2013scalable}, MANETs~\cite{wang2017trust}, and P2P~\cite{vasilomanolakis2017trust}. But, they take as basis previous interactions for the decision-making in handling data. Few studies turn to situations where previous interactions are unknown (zero-knowledge~\cite{feige1988zero}), in which there is only information from current interactions. Therefore, these solutions are not suitable for dynamic and sparse urban environments, as they assume the existence of a network infrastructure before data dissemination.

The ubiquitous presence of smartphones in people’s lives supports direct communication between them by means of technologies as bluetooth and WiFi~\cite{inbook}. Smartphones serve as gateways once they have multiple network interfaces, interconnecting users medical devices to the Internet~\cite{wood2017cleartext} to disseminate their health data. However, data dissemination must occur under control in networks to deliver data to the correct entities at the appropriate time~\cite{bujari2012survey}. In this context, the social aspects of the devices owners and from their relationships enable grouping network devices into communities~\cite{bao2013scalable}. Furthermore, some works employ trust as a criterion to control data dissemination~\cite{al2017trust}. Although suitable for unstructured networks, these works use reputation and recommendation models to evaluate trust, which in general are techniques dependent on the past interactions between devices.

This article presents the STEALTH (\textbf{S}ocial \textbf{T}rust-Based \textbf{HEALTH} Information Dissemination Control) system to disseminate personal sensitive health data in a controlled manner to achieve a minimum level of data confidentiality in dynamic wireless local area networks. It forms communities by grouping devices with common interests to create relationships of trust with each other. As long as the devices belong to a community and they are in the same wireless coverage range, they maintain connectivity with each other and their interactions occur between devices inside the community~\cite{bao2013scalable}. STEALTH considers the social aspects of devices owners and their relationships to measure trust of the neighbors. Under urgent situations, the system disseminates sensitive data controlling the process by choosing the appropriate device and user to receive them.

We evaluate STEALTH in the NS-3 simulator~\cite{NS3simulator2018} and we analyze its dependability to disseminate sensitive health data through evaluation metrics. To the best of our knowledge, this is the first work aimed at disseminating health data in dynamic urban environments, outside conventional environments (e.g., hospital or clinics). The results show that STEALTH reaches 97.14\% in data dissemination reliability and a maximum latency of \mbox{170 ms}, while data availability achieves 100\% under urgencies.

This article proceeds as follows. Section~\ref{sec:related} reviews the related works. Section~\ref{sec:background} introduces trust models in networks. Section~\ref{sec:systemModel} details the proposed system. Section~\ref{sec:eval} describes the evaluation methodology and results. Finally, Section~\ref{sec:conclusion} concludes the article and presents future directions.

\vspace{-0.3cm}
\section{Related Work}\label{sec:related}

Effective and fast data dissemination is a challenge~\cite{chuang2014time}. It comes from issues like dissemination frequency, dissemination location, and the nature of the content to be disseminated~\cite{sudhindra2014},~\cite{umarani2013survey}. Data dissemination also suffers from link losses, eavesdropping, and devices mobility~\cite{wallgren2013routing}. Considering these issues, one may define an appropriate strategy for data dissemination at the right time and to the right person~\cite{bujari2012survey}. Hence, there are different strategies in the literature to address these issues. The applied strategies comprise sending data in either an occasional or periodic fashion. Also, they send data from a source to a specific destination, or from a source to multiple destinations simultaneously.

Data dissemination raises security and safety concerns, particularly, if the information is sensitive and vulnerable to privacy attacks. The commonly applied techniques to data dissemination are broadcasting data~\cite{liu2018neighbor}, multi-hopping  transmission~\cite{daneels2017real}, and data replication~\cite{gonizzi2015data}. In dynamic scenarios (when there is a intense device mobility) , broadcasting data is an adequate solution~\cite{bujari2012survey}. The authors in~\cite{liu2018neighbor} proposed a protocol to disseminate data in mobile IoT networks. This protocol combines the neighborhood knowledge from nodes and adapts the connectivity factor to calculate the probability to determine whether a packet should be broadcast to other nodes or discarded, to prevent redundant packet broadcast. Although it is an effective strategy to data dissemination, it can cause channel collisions and compromise sensitive data privacy, as nodes do not control whom effective is receiving the disseminated data.

Multi-hopping data transmission can be applied in dynamic and infrastructure-free environments. In~\cite{daneels2017real}, the authors presented a general-purpose IoT platform based on a combination of Low-power Wireless Personal Area Network (LoWPAN) and multi-hop Wireless Sensor Network (WSN) technology. This platform provides multi-hop long-range connectivity between sensors and data sinks for real-time sensor data dissemination and analytics. This solution disseminates data in real-time, but it relies on the support of known devices to implement a multi-hop operation. Another strategy employed in data dissemination is data replication. In~\cite{gonizzi2015data}, the authors proposed a mechanism to deploy large scale observation systems in remote areas, when there is not a permanent connection with the Internet. The mechanism employs replication and distributed storage techniques to increase the amount of data stored within wireless sensor networks and to reduce the probability of data loss. This strategy addresses the infrastructure issue, but it does not meet the low latency requirements for sensitive data dissemination.

Trust techniques contribute to ensure safety and security in data dissemination, but there are challenges investigated by several works from the literature~\cite{cho2015survey}. Reputation~\cite{truong2017toward} and recommendation~\cite{al2017trust} models are techniques commonly applied, as well as communities of interest (CoI)~\cite{bao2013scalable}, to evaluate the trust level of devices in networks. In general, recommendation and reputation techniques lean on previous interactions of the devices, hence their behavior can be characterized over time. Few works focus on {\it zero-knowledge} environments, where maintaining interaction records is not always feasible due to the dynamic of environments, devices mobility and, sometimes, the constrained computational resources for processing and storing information, like in IoT. Dynamic interactions of devices over time are briefly useful.

In~\cite{al2017trust}, the authors presented a protocol based on recommendation techniques and sharing information between health devices in IoT for decision-making on access to specific sites. The protocol evaluates environments in which devices owners have been at a particular time and place, and it builds a database for the future decision-making process. The recommendation techniques incorporate characteristics of social relationships from device owners to manage the access to environments. However, they require many interactions between devices, what is not always possible in dynamic network environments. Therefore, those techniques can inhibit the  construction of database, making the decision-making process unfeasible.

In~\cite{truong2017toward}, the authors presented the Reputation, Experience and Knowledge (REK) approach, whose goal is to assess trust based on multi-dimensional aspects: \textit{reputation}, \textit{experience} and \textit{knowledge}.  \textit{Reputation} consists in a public opinion about who is evaluated. \textit{Experience} takes into account previous interactions with who is evaluated. \textit{Knowledge} corresponds to the understandings about who is evaluated. Although reputation allows choosing specific devices to perform critical tasks, it demands knowledge of devices interactions over time. Recommendations and experience face the same issue. Also, measuring subjective indicators, such as knowledge and experience, is a challenging~\cite{cho2015survey}.

In~\cite{bao2013scalable}, the authors proposed a protocol to manage trust in social IoT environments, where conditions are dynamically changing (e.g., increasing misbehaving node population/activity, changeable behavior, rapid membership changes, and interaction pattern changes). The protocol establishes communities clustering devices based on recommendations and the relational trust attributes from the social relationships of device owners, such as honesty and cooperation. Network devices can participate in a cluster or leave it any time. However, as recommendations rely on past interactions, the protocol is unsuitable to dynamic environments with eventual interactions.

Although all the above works offer important contributions, they do not make it possible to disseminate sensitive data outside traditional care institutions. Existing solutions employ techniques (e.g., reputation~\cite{truong2017toward}, recommendation~\cite{al2017trust}, experience, and knowledge~\cite{truong2017toward}) that depend on past interactions of the network devices. Hence, those techniques inhibit to work under~\textit{Zero-Knowledge} conditions~\cite{feige1988zero}. Most importantly, data dissemination by broadcasting data~\cite{liu2018neighbor}, multi-hopping  transmission~\cite{daneels2017real}, and data replication~\cite{gonizzi2015data} faces issues like devices mobility and sparse environments. These issues demand controlled data dissemination, avoiding unauthorized access to data. STEALTH overcomes these shortcomings.

\vspace{-0.3cm}
\section{Background and Overview} \label{sec:background}

Ensuring robustness in the dissemination of sensitive data on IoT requires a level of trust among the existing devices~\cite{guo2017survey}, that can be obtained from social aspects of their owners (i.e., characteristics of people and social relationships). The application domain defines the social aspects to be employed (e.g., emotional, logical, and relational trust). In dynamic environments, where there is no history of previous device interactions (i.e., \textit{Zero-Knowledge} conditions~\cite{feige1988zero}), proposals for evaluating trust are still incipient.

A range of trust techniques has supported solutions in different networks, like Mobile Ad Hoc Networks (MANETs)~\cite{wang2017trust}; Peer-to-Peer (P2P)~\cite{vasilomanolakis2017trust, zuo2016survey}; and, recently, the Internet of Things (IoT). In IoT, one observes trust techniques through centralized ~\cite{al2017trust, cervantes2015detection, son2017adaptive}, and distributed approaches~\cite{bao2012dynamic, bao2012trust, bao2013scalable, nguyen2016challenge, oh2018unified, truong2017toward}. A set of techniques stands out in the trust composition, such as message exchange, communities of interest, and recommendation. Among the attributes for trust evaluation, there is the number of messages received and forwarded, interactions, and activities. Communities group devices based on some criteria (i.e., common characteristics or interests).

The trust assessment of network devices follows different approaches. The number of data writing and forwards within predicted limits may indicate a trustworthy device. By exchanging specific messages, the authors have demonstrated the correct functioning of a system or mechanism~\cite{vasilomanolakis2017trust}. Few evaluation techniques take as basis social relationships among devices owners, observing information, such as user profiles, location, or interests. Information from social networks, like Facebook, Linkedin, and Foursquare, for example, improves the evaluation of the relationships between network devices; meanwhile, they restrict access to information to only trusted ones~\cite{zuo2016survey}.

Reputation techniques are commonly part of a trust assessment process of IoT devices. This may happen through centralized~\cite{al2017trust, cervantes2015detection, son2017adaptive}, or distributed approaches~\cite{bao2013scalable, bao2012dynamic, bao2012trust, nguyen2016challenge, oh2018unified, truong2017toward}. Although centralized approaches for IoT are not commonly found in the literature, there are any works available. These solutions usually assess trust through reputation techniques and the number of messages received and forwarded. These approaches do not meet all IoT configurations, mainly in dynamic environments (i.e., where devices are highly mobile and eventually participate in the network). Distributed approaches suit better  to IoT, and, generally, the solutions associate reputation with other trust techniques, like recommendation. This process improves efficiency and effectiveness. Intrusion detection systems employ reputation techniques~\cite{cervantes2015detection}. They combine reputation with watchdog strategies and trust to group devices and handle their density and mobility. Therefore, they can detect attacks in routing, and prevent, identify, and isolate their effects on the network~\cite{cervantes2015detection}. In such cases, they calculate the reputation of a network device from the relation between the number of receiving and transmitting data. They classify the device behavior based on that relation. Devices propagate their reputation to others in the network, allowing them to also calculate trust.

Recommendations are indirect observations and allow devices to share information, for instance, about particular environments (e.g., temperature, humidity, and time)~\cite{al2017trust}. This procedure helps in decision making, such as deciding whether people can access some locations. Recommendations make feasible to compute the reputation of network devices, taking into account direct observations that other nodes make about it. Besides that, the choice of recommendations to the detriment of older information contributes to the increasing convergence speed of trust~\cite{bao2012trust}. The relationship between device owners at IoT enables grouping devices to establish clusters or communities, given social aspects such as honesty and cooperation~\cite{bao2012dynamic, bao2012trust, bao2013scalable}. Device interactions inside communities are more intense than between distinct communities, where data traffic reduces to some extent. In this context, devices perform trust assessment within the cluster they belong to. Besides, CoIs allow coping with network scalability. 

\vspace{-0.3cm}

\section{System Model} \label{sec:systemModel}

This section provides an overview of the network model and the STEALTH architecture. Next, we illustrate an example of the operating STEALTH.

\vspace{-0.2cm}

\subsection{Network Model} \label{sec:models}

\begin{figure}[!h]
\centering
\includegraphics[width=.45\textwidth]{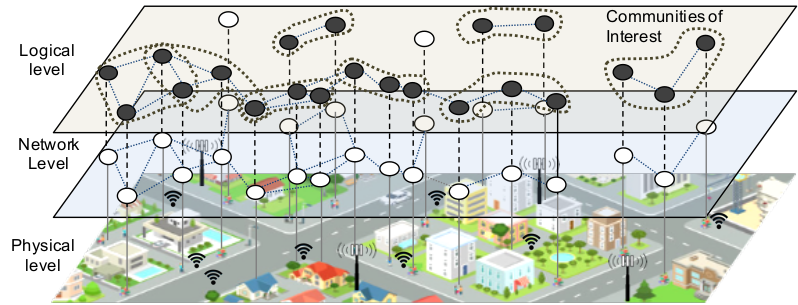}
\caption{Network model and sensitive data dissemination}
\label{fig:ModeloRede2}
\end{figure}

STEALTH relies on social aspects and relationships from the device owners to create local networks over time and to maintain communities of interest, as depicted in Fig.~\ref{fig:ModeloRede2}. Under a critical event leading to an emergency, STEALTH disseminates sensitive data to appropriate receivers who are physically close, taking as basis the receiver competence. Hence, STEALTH supports emergency care to a person.

\mbox{STEALTH} carries out on a set of portable devices (nodes), denoted by $D = \{d_1,d_2,d_3,..., d_j\}$, where $ d_j \in D $, interconnected in a wireless communication network. These nodes have processing and communication resources to group nodes and disseminate data. Each node possesses a unique identifier ($Id$) to identify it over time. Each device leverages the competence (skill) and interests from its owner or user, as attributes for trust calculation in STEALTH. Hence, in a skill set $S = \{s_1,s_2,s_3,...,s_k\}$, $|S| \neq 0$, a competence level $s_m$, such that $ s_m \in S $, is a value in the range 0 to 1, represents a skill or knowledge from the owner or user of a device $d_w$, where $d_w \in D$, in a particular field of activity (e.g., doctor, police officer, other). Each node also holds a set of interests $I_n = \{i_1,i_2,i_3,...,i_z\}$, where $|I_n| \neq 0$ and $I_n \subset I$, and $I$ is the set of all interests. An interest is a hobby, taste or preference (e.g., music, health, other). 

Nodes are grouped by common interests and form communities over a given period of time. A community $C$ is a set of distinct tuples $\langle$node, period, interest$\rangle$, where $C = \{\langle d_1,P_l,i_z \rangle, \langle d_2,P_l,i_z \rangle,...,\langle d_n, P_l,i_z \rangle\}$ and $P_l = ((t_ {s0},t_ {e0}),(t_ {s1},t_ {e1}),...,(t_{sl},t_{el}))$, with $ t_ {s *} \leq t_{e*} $.\footnote{Definition adapted from the concept of dynamic communities proposed by~\cite{coscia2011} and revised by~\cite{rossetti2018community}} The efficiency of applying node interests as a criterion for forming communities is associated with their closeness, while competence are effective inside each community. By simplicity, we assume that disconnected or intermittent failing nodes do not act on the network. Also, connected nodes behave in an honest manner, disregarding attacks on system health.

\subsection{Dissemination Control to Personal Sensitive Data} \label{sec:stealth}

\begin{figure}[!ht]
\centering
\includegraphics[width=.45\textwidth]{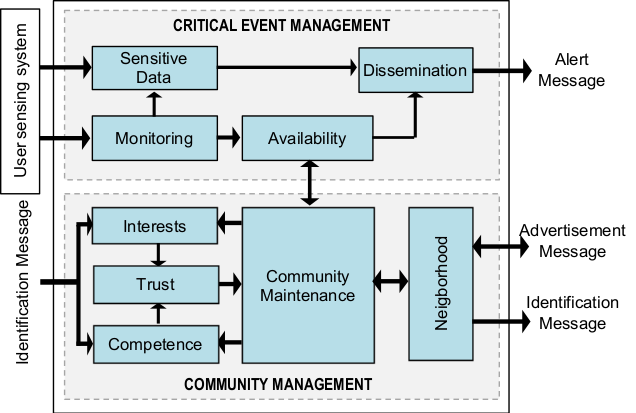}
\caption{STEALTH Architecture}
\label{fig:ArchitectureStealth}
\vspace{-0.5cm}
\end{figure}

Fig.~\ref{fig:ArchitectureStealth} illustrates the STEALTH (\textbf{S}ocial \textbf{T}rust-Based \mbox{\textbf{HEALTH}} Information Dissemination Control system) architecture that comprises two main modules: the {\bf Community Management} module (CMM) and the {\bf Critical Event Management} module (CEM). The first module is responsible for creating and updating the communities of interest established over time from the interaction among people. The second module is responsible for verifying and disseminating the sensitive data of a person in an emergency, The next subsections describe each module.

\subsubsection{Community Management Module}

This module measures the trust level of nearby devices and includes them in a community. A device joins a community depending on its health interest and upon receiving the node \textit{identification message} with its \textit{Id}, interests, and competence. CMM is also responsible for identifying the node to a neighbor that is searching for neighboring nodes to form their communities. It comprises five components: \mbox{\textit{Neighborhood}}, \mbox{\textit{Interests}}, \mbox{\textit{Competence}}, \mbox{\textit{Trust}} and \mbox{\textit{Community Maintenance}}. The \mbox{\textit{Neighborhood}} component searches for neighboring nodes. It sends an \textit{identification message} to other nodes that are searching for neighbors to identify its neighborhood. The \mbox{\textit{Interests}} component analyzes the interests of neighboring nodes when receiving them, it identifies common interests to group nodes and form communities of interest. The \mbox{\textit{Competence}} component deals with the competence of the neighboring nodes when receiving it to know their health competence level. Based on their interests, the \mbox{\textit{Trust}} component measures the trust level of neighboring nodes when receiving their interests and competence. It verifies on their health community the neighbor with the highest health competence level. Finally, the \mbox{\textit{Community Maintenance}} component coordinates the creation, extinction, and modification of CoIs, from nodes interactions. It ensures the communities of interest following the evolution of local networks established over time.

\SetKwComment{Comment}{$\triangleright$\ }{}

\SetAlFnt{\footnotesize}
\begin{algorithm}[!t]
{
{
\textbf{for each node $d \in D$ do} \\
\hsep \textbf{procedure} $\textsc{SearchNeighbors}$ \\
\hsep \hsep \textbf{while} (true) \textbf{do}\\
\hsep \hsep \hsep $NL \leftarrow 0$ \\
\hsep \hsep \hsep $SendAnnounce(\;)$ \\
\hsep \hsep \hsep $WaitInterval(\;)$ \\
\hsep \hsep \textbf{end while}\\
\hsep \textbf{end procedure}\\

\BlankLine

\hsep \textbf{procedure} \textsc{ReceiveAnnounce}$  (\;)$ \\
\hsep \hsep $nskill \leftarrow GetSkill (\;)$ \\
\hsep \hsep $ninterest \leftarrow GetInterests (\;)$ \\
\hsep \hsep $AnswerAnnounce(id,\;nskill,\;ninterest)$ \\
\hsep \textbf{end procedure} \\

\BlankLine

\hsep \textbf{procedure} \textsc{ReceiveAnswer} $(id,\;nskill,\;ninterests)$ \\
\hsep \hsep \textbf{if} ($CommonInterests(ninterests)$ AND \\
\hsep \hsep $HealthInterest(ninterests)$) \\
\hsep \hsep \hsep $ntrust \leftarrow EvaluateNeighborTrust (nskill, \; ninterests)$ \\
\hsep \hsep \hsep $NL \leftarrow RegNeighbor (id, \; nskill, \; ninterests, \; ntrust)$ \\
\hsep \hsep \textbf{end if}\\
\hsep \textbf{end procedure}\\

\BlankLine

\hsep \textbf{procedure} \textsc{EvaluateNeighborTrust} $(nskill, \; ninterests)$ \\
\hsep \hsep $skilltrust \leftarrow GetSkillTrust (skill, SkillsTaxonomy)$ \\
\hsep \hsep $ncinterests \leftarrow GetNCInterests (interests)$ \\
\hsep \hsep $nninterests \leftarrow GetNNInterests (\;)$ \\
\hsep \hsep $itrust \leftarrow ncinterests \; / \; nninterests$ \\
\hsep \hsep \textbf{return}  $(skilltrust \; + \; intereststrust) \; / \; 2$ \\
\hsep \textbf{end procedure}\\
\caption{Community Management}
\label{alg:coi}}}
\end{algorithm}

Network nodes start operating in an isolated way and, as they move, find other nodes, and establish communities of interest with those that are health-interested. As described in Algorithm~\ref{alg:coi}, each node periodically initializes its neighbor list (\textit {l}.3), announces its presence by broadcasting messages (\textit{l}.4) searching for neighboring nodes and wait for a time interval to a new announcement (\textit{l}.5). When a neighbor node takes notice that a node announces its presence (\textit{l}.8), it forwards to this announcing node an \textit{identification message} (\textit{l}.11). Upon receiving this message, the announcing node verifies whether they are both health-interested (\textit{l}.14). If they are, the announcing node measures the trust level of the neighboring node through \textit{EvaluateNeighborTrust} (\textit{l}.15) and it includes the neighboring node into its neighbor list (\textit{l}.16) inside its health community. This takes into account the trust level of the neighboring node, its competence (\textit{l}.20) and the common interests with the announcing node~(\textit{l}.21-23).

\subsubsection*{Competence Classification}\label{sec:classifCompetencia}

\begin{figure}[!ht]
\centering
\includegraphics[width=.51\textwidth]{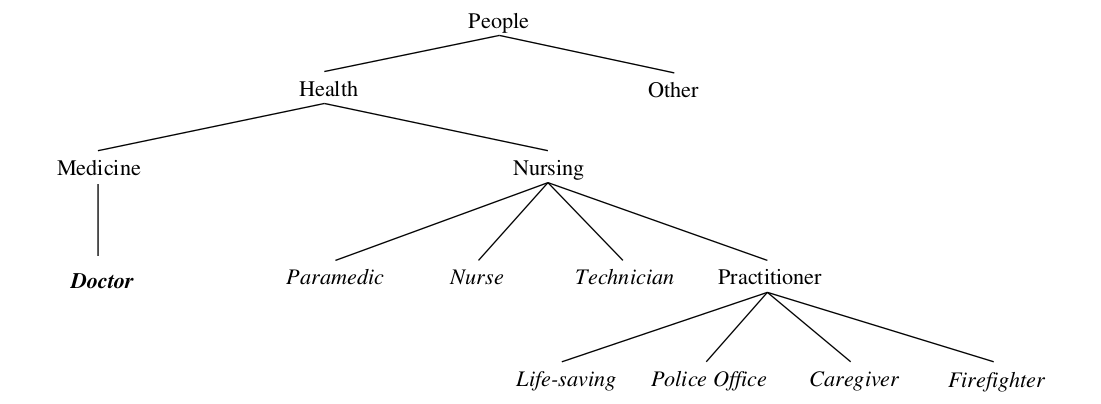}
\caption{Healthcare skill taxonomy}
\label{fig:taxSkills}
\end{figure}

Every device user possesses certain abilities to perform the daily activities, which are competences obtained from professions, skills, or hobbies, for example. STEALTH takes into account competences to disseminate sensitive data to the right person (i.e., who is health-skilled). Therefore, a set of health competences is hierarchically organized employing the level of each profession knowledge as a criterion. This organization gives rise to a skill taxonomy ($S_T$), depicted in Fig.~\ref{fig:taxSkills}, which distributes health competences of each profession in levels, as proposed by~\cite{carminati2016detection}. 

STEALTH follows an extension of this taxonomy with other competences, according to the current need. It classifies nodes based on the health knowledge of the device owners. It organises devices related to people with healthcare knowledge into two distinct areas - medicine and nursing. In the branch of medicine, there are doctors. In the field of nursing, there are nurses and several other professionals, as well as a class of professionals with reduced health skills, here called practitioners. This group encompasses caregivers, police officers, firefighters, and other professionals trained to provide first aid.

\vspace{-0.5cm}

\vspace{-0.3cm}

\begin{figure}[!t]
\centering
\includegraphics[width=.2\textwidth]{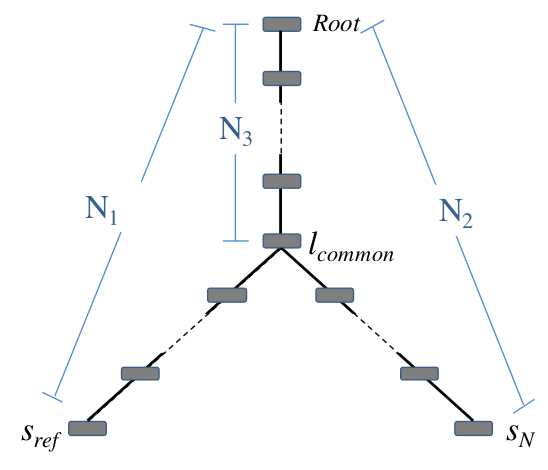}
\caption{Measure of similarity between skills}
\label{fig:medidaSimilaridade}
\vspace{-0.5cm}
\end{figure}

STEALTH evaluates the similarity of people competence employing a skill taxonomy based on~\cite{wu1994verbs, mohammad2012distributional} and depicted in Fig.~\ref{fig:taxSkills}. Fig.~\ref{fig:medidaSimilaridade} illustrates the verification of the similarity of a competence. In this work, the reference competence ($ s_{ref}$) is a doctor, because this professional has the highest healthcare knowledge. STEALTH calculates $Sim_s$ through~(\ref{eq:competSim}), where $N_3$ corresponds to the number of levels from the common level ($l_{common}$) closest to the neighbor competence and the reference competence, to the root of the taxonomy ($Root$). The $N_2$ is the number of levels from the neighbor competence ($s_{Neigh}$) to the root of the taxonomy ($Root$), and $N_1$ is equal to the number of levels from the reference competence (\textit{doctor}) to the root of the taxonomy ($Root$). The values of $Sim_s$ vary in the range $[0,1]$, as showed in ~(\ref{eq:valuesSim}).

\begin{equation}
\footnotesize
Sim_s = \frac {2 \times N_3}{N_1 + N_2}
\label{eq:competSim}
\end{equation}

\begin{equation}
\footnotesize
Sim_s = \left \{
\begin{array}{cl} 
0,	& if \; s \; \in \{other\} \\
]0,1[,	& if \; s \; \notin \{other, doctor\} \\
1, 	& if \; s \; \in \{doctor\}
\end{array}
\right.
\label{eq:valuesSim}
\end{equation}

Assuming the necessity to evaluate the similarity between \textit{life-saving} ($Sim_s$) and \textit{doctor} competences, we follow the taxonomy in Fig.~\ref{fig:taxSkills}. The competence of a doctor is three levels up to the root of the taxonomy and corresponds to $N_1$. The \textit{life-saving} competence is four levels up to the root of the taxonomy and equals to $N_2$. Finally, STEALTH obtains the distance of the common level from the assessed and reference competences (\textit{health}) to the root of the taxonomy, $N_3 = 1$. By means of~(\ref{eq:competSim}), STEALTH obtains $Sim_s = 0.4$, which is the similarity of the \textit{life-saving} to \textit{doctor} competence.

\subsubsection*{Trust Measurement}

STEALTH measures trust level for the nodes based on social aspects of their devices owners: an individual - \textit{Competence} - and a relational one - \textit{Similarity}. Competence is a skill perceived in a node, inherited from the device owner to perform a task~\cite{cho2015survey} (e.g., a profession, a hobby). Trust is a variable value, and it increases as the healthcare competence of the assessed node resembles the \textit{doctor} competence. Similarity is related to common interests that the evaluating node and the evaluated one possess. Therefore, trust value increases as the number of common interests increases. A node evaluates the trust of other nodes only if they are health-interested. Hence, the measurement occurs whenever the evaluated node possesses at least health interest, which implies a minimum trust value always greater than 0.

One consider a node $x$ encounters a node $y$ and measures its trust level about the common interests between them, $T_{xy}^I$. It is a ratio between their common interests, $I_x \cap I_y$, and the interests of the evaluating node itself, $I_x$. $T_{xy}^I$ is obtained through~(\ref{eq:communityTrust}), which is based on~\cite{bao2012trust}. The values of $T_{xy}^{I}$ vary in the range $[0,1]$, as showed in~(\ref{eq:valuesTCoI}). This trust measurement occurs whether node $y$ is health-interested.

\begin{equation}
\footnotesize
T_{xy}^{I} = \frac {|I_x \cap I_y|}{|I_x|}
\label{eq:communityTrust}
\end{equation}
\quad
\begin{equation}
\footnotesize
T_{xy}^{I} = \left \{
\begin{array}{cl}
0, & if  \; \; I_y \not\supset \{health\} \\
]0,1[,	& if \; \; I_x \cap I_y \neq 0 , \; \; I_x \neq I_y \; and \; \{health\} \subset I_x \cap I_y\\
1, 	& if \; \; I_x = I_y \; \; and \; \; \{health\} \subset I_x \cap I_y
\end{array}
\right.
\label{eq:valuesTCoI}
\end{equation}

Checking the similarity of node $y$ competence with that of a doctor enables computing trust about node $y$ competence, $T_{xy}^{Skill}$. STEALTH deems the competence of a doctor as the highest in health and the calculation of $T_{xy}^{Skill}$ is done on a skills taxonomy ($S_T$) presented in Fig.~\ref{fig:taxSkills}, based on~\cite{carminati2016detection,mohammad2012distributional}. Therefore, $T_{xy}^{Skill}$ equals the distance ($D_{ST}$) from node $y$ competence ($s_y$) relative to the health competence within that taxonomy, and is obtained by~(\ref{eq:SkillTrust}). In this work, we assume a distance function $D_{ST}(S)$ that receives as input a competence $S$ from a evaluated node and returns a value in the range of $[0,1]$. This value indicates the closeness of informed competence to doctor competence in the taxonomy $S_T$. The distance function $D_{ST}$ is based on the measure established by~\cite{wu1994verbs} and revised by~\cite{resnik1999semantic}. The possible values of $T_{xy}^{Skill}$ vary in the range $[0,1]$, as in~(\ref{eq:valuesTSkill}).

{
\begin{equation}
\footnotesize
T_{xy}^{Skill} = Sim_y
\label{eq:SkillTrust}
\end{equation}

\quad

\begin{equation}
\footnotesize
T_{xy}^{Skill} = \left \{
\begin{array}{cl} 
0,	& if \; s_y \; \in \{other\} \\
]0,1[,	& if \; s_y \; \notin \{other, doctor\} \\
1, 	& if \; s_y \; \in \{doctor\}
\end{array}
\right.
\label{eq:valuesTSkill}
\end{equation}
}

Finally, the trust level of node $x$ over node $y$ ($T_{xy}$) is $0$, whether they are not both health-interested. Otherwise, $T_{xy}$ corresponds to the sum of trust related to their common interests, $T_{xy}^{I}$, with that derived from node $y$ competence ($T_{xy}^{Skill}$), as in~(\ref{eq:totalTrust}). When $T^I_{xy}>0$, the values of $T_{xy}$ vary in the range $]0,1]$, depending on the values of $T_{xy}^I$ and $T_{xy}^{Skill}$, as showed in~(\ref{eq:valuesTCoI}) and~(\ref{eq:valuesTSkill}), respectively.

\begin{equation}
\footnotesize
T_{xy} = \frac{T_{xy}^{I} + T_{xy}^{Skill}}{2}
\label{eq:totalTrust}
\end{equation}

For example, considering a node $x$ that assesses the trust level of a node $y$, whose competence is \textit{caregiver}, and both hold a single interest, \textit{health}. Employing the skills taxonomy ($S_T$) presented in Fig.~\ref{fig:taxSkills} and the interest described, $Sim_s(caregiver)$ will hold a value of 0.28 (i.e., $T_{xy}^{Skill} = $ 0.28). $T_{xy}^{I}$ holds value 1, as it's calculated by~(\ref{eq:communityTrust}), since nodes are health-interested only. Hence, through~(\ref{eq:totalTrust}), $T_{xy} = 0.64$.

\subsubsection{Critical Event Management Module}

In this module, the \textit{Monitoring} component verifies a person health condition upon receiving her health status from her sensing system. The medical device, carried by a person, is responsible for identifying a critical event and reporting to STEALTH. The \textit{Sensitive Data} component obtains the person sensitive data in an emergency and ensures its dissemination only under these conditions. The \textit{Availability} component verifies the appropriate device to disseminate data, ensuring that it is the one with the highest health competence. The \textit{Dissemination} component coordinates the sensitive data dissemination upon receiving them and the identification of the appropriate person. This dissemination occurs through alert messages sent only to people who belong to the health community of the node and allowing for its health competence.

\begin{algorithm}[!t]
{
{
\textbf{for each node $d \in D$ do} \\
\hsep \textbf{procedure} \textsc{HandleEmergencyEvent}$  (\;)$ \\
\hsep \hsep $neighid \leftarrow GetHigherScoreNeighbor (\;)$ \\
\hsep \hsep $neighskill \leftarrow GetNeighborSkill \; (neighid)$ \\
\hsep \hsep $criticaldata \leftarrow GetCriticalData \; (neighskill)$ \\ 
\hsep \hsep $SendAlert (neighid,\;criticaldata)$ \\
\hsep \hsep $SendStopAnnounce(\;)$ \\
\hsep \textbf{end procedure}\\

\hsep \textbf{procedure} \textsc{ReceiveAlert} ($id, \; criticaldata$) \\
\hsep \hsep $SendAckAlert(\;)$ \\
\hsep \textbf{end procedure} \\

\hsep \textbf{procedure} \textsc{ReceiveStopAnnouce} ($Id$) \\
\hsep \hsep $NL \leftarrow RemoveNeighbor(Id)$\\
\hsep \textbf{end procedure} \\

\caption{Critical Event Management}
\label{alg:emerg}}}
\end{algorithm}

Nodes belonging to a health community support the nodes that represent a person in an emergency, as described in Algorithm~\ref{alg:emerg}. When a critical event occurs with a particular node (i.e., a person entered in an emergency), it verifies the neighbor node with the highest trust level (\textit{l}.2) and obtains the appropriate sensitive data from the person in an emergency (\textit{l}.3-4). Next, it sends an alert message to the selected node (\textit{l}.5) with its sensitive data. Besides, it announces by broadcast the interruption of its operation (\textit{l}.6). Upon receiving an alert message, the node acknowledges it (\textit{l}.9). When a node notices that another node announces that is interrupting the operation, it removes the announcer node from its neighbor list (\textit {l}.11). Therefore, this reaction prevents a node in an emergency from being selected to receive sensitive data from other nodes.

\subsection{Operation} \label{sec:operation}

This section illustrates the operation of the \mbox{STEALTH} system in an urban environment and demonstrates its contribution to the controlled dissemination of sensitive data in an emergency, hence the user can receive a first aid. The illustrative scenario lies in a metropolitan area where six people walk down the streets: a nurse, a patient, an executive, a police officer, a fireman, and a doctor. Each one has a profession or ability to perform specific tasks daily. While the doctor is the most health-skilled, the police officer, for instance, has some health knowledge to provide first aid. The patient eventually needs emergency care.

All these people are health-interested and do not maintain relationships with each other. The nurse, the police officer, the fireman, and the doctor are health-interested because of their profession. The executive, for example, is health-interested to help people in need. However, the patient is health-interested to receive some aid. All those people carry mobile devices (i.e., smartphones) to connect to networks. \mbox{STEALTH} runs on these smartphones and it is set to operate. The patient carries a medical device close or on the body to analyze the blood pressure, for instance, and report to an application installed on the smartphone. The application reports the measured blood pressure values and their normality for the patient.

\begin{figure}[!t]
\centering
\includegraphics[width=.33\textwidth]{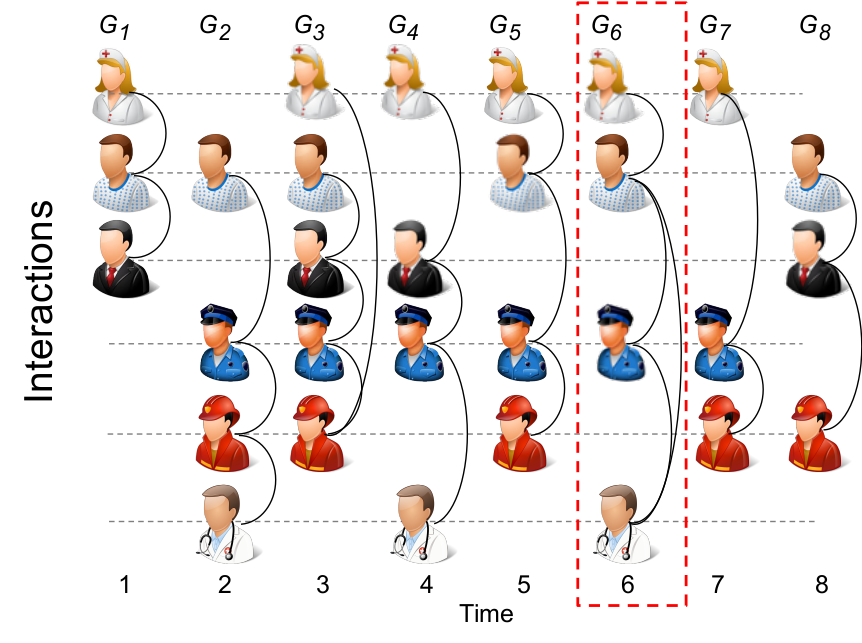}
~~
\raisebox{13mm}{\includegraphics[width=.11\textwidth]{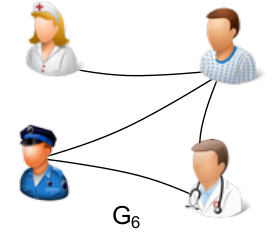}}
\vspace{-0.4cm}
\caption{Interactions over time}
\label{fig:interacoesnotempo}
\vspace{-0.1cm}
\end{figure}

\begin{table}[!htb]
\setlength{\extrarowheight}{1.0pt}
\centering
{\footnotesize
\caption{Trust Measurement}
\vspace{0.1cm}
\label{tab:exemploConfianca2}
\begin{tabular}{l|ccc}
\hlineB{2}
\multirow{2}{*}{\textbf{Trust}}&\multicolumn{3}{c}{\textbf{Competence}} \\ \cline{2-4} 
&doctor&Nurse&Police Officer  \\ \hline
\textbf{$T^{Skill}$}&1&0.33&0.28  \\
\textbf{$T^{CoI}$}&1&1&1  \\
\textbf{$Trust$}&1&0.66&0.64  \\
\hlineB{2}
\end{tabular}}
\vspace{-0.2cm}
\end{table}

People interactions change over time ($t = \{1,2,...,8\}$) due to their mobility, as illustrated in Fig.~\ref{fig:interacoesnotempo} left. Their devices establish ad hoc networks for data exchange. At $t_6$, the device of the patient and his smartphone interact with other users, as represented by graph $G_6$ (Fig.~\ref{fig:interacoesnotempo} right) and each device forms its health community. The device of the patient measures the trust level of neighboring nodes and inserts them into its neighbor list with trust values displayed in Table~\ref{tab:exemploConfianca2}. On the assumption that the patient enters an emergency at $t_6$ (i.e., a critical event occurs at $t_6$), STEALTH running on the smartphone of the patient verifies in his health community that the doctor is the person with the highest trust, and disseminates the sensitive data to him.

\vspace{-0.3cm}
\section{Performance Evaluation}\label{sec:eval}

This section presents the performance evaluation for availability and reliability in the dissemination of people sensitive data in an emergency. The next subsections describe the simulation settings, scenarios and results. 

\vspace{-0.2cm}
\subsection{Simulation settings}\label{sec:settings}

We employ simulations to evaluate STEALTH using the NS-3 simulator, version 3.28. The environment has included a Dell Inspiron with Intel(R) Core(TM) i5-4460S CPU@2.90GHz 64 bits, 8GB RAM. A VM VirtualBox, version 5.2.18, r124319, supports a Debian operating system, version 9.1, to execute NS-3. Table~\ref{tab:simSettings} shows the main simulation settings. We have conducted analysis in a scenario of 100 mobile devices (nodes), following a user mobility model in an urban environment. These users carry a portable equipment (i.e., smartphone). They had been deployed in a 400 m x 430 m area. Users have walked in this area with speeds ranging from 0.5 m/s to 2.0 m/s. 

\begin{table}[!t]
\setlength{\extrarowheight}{1.0pt}
\centering
{ 
\small 
\caption{Simulation settings}
\vspace{0.1cm}
\label{tab:simSettings}
\scriptsize
\begin{tabular}{|l|l|}
\hline
\textbf{Parameters} & \textbf{Values} \bigstrut \\ \hline
\# of Nodes & 100 \\
Nodes evaluated & 37, 52, 69 and 70 \\
Simulation time & 900 sec\\
Emergency time & 300 sec and 485 sec \\
Nodes speed & 0.5 m/s to 2.0 m/s \\
Area & 400 m x 430 m \\
Standard &  IEEE 802.11a \\
Package type & UDP \\
Transmission radius & 50 m \\
\hline
\end{tabular}}
\vspace{-0.4cm}
\end{table}

Users have followed the mobility model introduced in ~\cite{helgason2014opportunistic},~\cite{helgason2010does}. This is a realistic model implemented in Legion Studio, a pedestrian simulator used for designing large public spaces. The mobility is based on analytical models, which allow mimic aspects of an individual movement including personal preferences, surrounding awareness, and perception of behaviors~\cite{legionStudio}. The mobility patterns follow the \textit{least effort} principle, where, just like in reality, each entity tries to minimize the dissatisfaction before choosing its next move. The employed scenario models the \"Ostermalm area of central Stockholm and consists of a grid of interconnected streets, where each street is 2 m wide and lengths vary between 20 m and 200 m. There are 12 passages that connect the area to the outside world and one assumes that all streets have equal node arrival rates. Upon arriving at an intersection, nodes continue to move on the same street (if possible) with probability of 50\% or turn to other adjoining streets with equal probability. The mobility trace file contains a snapshot of the positions of all nodes every 0.6 sec. The model considers a Poisson arrival process due to the uniform and the truncated normal distributions resemble reality to a higher extent~\cite{helgason2010does}.

Nodes have established ad hoc networks through transmission applying the IEEE 802.11a standard and the UDP transport protocol. A  \mbox{50 m} radius transmission allows nodes creating communities to the extent that they move. Besides, we configure nodes randomly with social aspects at each simulation repetition. They keep a single competence and a set of interests, with a minimum of one and a maximum of five interests. Table~\ref{tab:aspectosAtribuidos} presents the distribution of the assigned social aspects to the nodes. We have extended NS-3 class \textit{node} to incorporate social attributes to nodes. The guidelines for running the application and its codes, which include STEALTH, the modified NS-3 \textit{node} class, and mobility \textit{traces} can be found at GitHub.\footnote{https://github.com/agnaldosb/stealth}

We label nodes in simulation from 1 to 100. As defined in each scenario, we have performed system behavior evaluation through four of them: 37, 52, 69, and 70. These nodes keep the same configuration in all carried-out simulation repetitions, while the other 97 nodes are randomly configured in each simulation repetition. The total simulation time is 900 s, and the selected nodes come into an emergency at \mbox{300 sec} and \mbox{485 sec} of a simulation repetition, according to the evaluated scenario. We assume that all nodes exhibit honest behavior, and there are security mechanisms for validating their identities and protecting data transmission. We also consider that people carry a device close to the body responsible for the identification of a critical event and to inform STEALTH. Results correspond to an average of 35 repetitions with a 95\% confidence interval.

\begin{table}[!t]
\setlength{\extrarowheight}{1.0pt}
\centering
{\scriptsize
\caption{Distribution of social aspects assigned to nodes}
\vspace{0.1cm}
\label{tab:aspectosAtribuidos}
\begin{tabular}{l|c||l|c}
\hlineB{2}
\textbf{Competence} & \textbf{\# of Nodes} & \textbf{Interest} & \textbf{\# of Nodes} \\ \hline
Doctor        & 10    &  Health    & 20    \\
Nurse       & 15 & Music   & 30 \\
Caregiver   & 20 & Tourism & 45 \\
Other       & 25 & Movies  & 60 \\
             &   & Books   & 15 \\
\hlineB{2}
\end{tabular}}
\vspace{-0.4cm}
\end{table}

\subsubsection{Evaluated scenarios}\label{sec:configEspecif}

We employ three distinct evaluation scenarios in STEALTH analysis, as detailed next. 

\textit{\textbf{SENACK: Single event, access to sensitive data, non acknowledgement}.} Three nodes (i.e., 37, 52, and 70) always behave in the same way at every repetition. They had been selected because they move throughout the total simulation time and travel the longest paths in the selected urban environment. We assign to those nodes the same competence in all repetitions - \textit{other} - and all possible interests - \textit{health, tourism, music, movies,} and \textit{books}. The critical events occur at the simulation time \mbox{300 sec}, which represents the moment a person enters in an emergency. At this time, nodes have moved intensively and often interact to each other. Therefore, there was always a neighborhood around. 

\textit{\textbf{SEACK: Single event, access to sensitive data, acknowledgement}.} It is similar to SENACK, but the node that receives sensitive data must acknowledge it. While this confirmation is not received, nodes in an emergency continue to look for neighboring nodes and keep their health community up to date. Nodes stop searching for new neighboring nodes and receiving new messages when it receives the acknowledgement or when there is no other neighboring node in its health community. Critical events have occured at \mbox{300 sec} of the simulation time.

\textit{\textbf{MEACK: Multiple events, access to sensitive data,  acknowledgement}.} MEACK is similar to SEACK, except for the nodes in an emergency disseminate their sensitive data and a service priority indicator. This indicator makes it possible decision-making process about the attending order that must be followed by the appropriate entity when receiving simultaneously sensitive data from multiples persons. Its an integer value varying from 1 to 4, where 1 indicates that a person needs the highest priority in health attending, while 4 is the lowest one. Thus, when multiple nodes are in an emergency and disseminate their sensitive data to a single node, this node verifies the service priority indicators received in order to acknowledges in the correct order. If the received indicators holds distinct priorities, the acknowledgement occurs in descending order of priority (i.e., from the highest to the lowest). Otherwise, the acknowledgement follow the sensitive data reception order. The nodes selected to meet this are 52, 69, and 70. A fourth node, 63, is responsible for receiving data from those nodes. We have chosen these nodes because they were close enough at the instant \mbox{485 sec} of the simulation to exchange data. We have assigned to nodes 52, 69, and 70 the same competence in all repetitions -~\textit{other}~- and all possible interests - \textit{health, tourism, music, movies,} and \textit{books}. Node 63 has also received all the same interests, but a different competence  - \textit{doctor}. Hence, it has achieved the highest trust level within the health communities of the neighboring nodes, increasing its chance of being selected to receive sensitive data.

\subsubsection{Metrics}\label{sec:metrics}

We employ specific evaluation metrics aiming to verify the network behavior and its performance. Therefore, we discuss the results about the urban pedestrian mobility behavior through the metric \textit{Average Number of Neighbors} ($N_{N}$). The evaluation of data availability provided by \mbox{STEALTH} takes into account the evolution of health communities over time and the \textit{Average Number of Health Communities} ($N_{C}$). The assessment of data reliability in the dissemination service follows the metrics: \textit{Hit Rate} ($HR$), \textit{Fault Rate} ($FR$), \textit{Average Time to Access Sensitive Data} ($AT$), and \textit{Hit Rate by Competence} ($HR_{Skill}$). Next, all these metrics are described.

\textbf{\textit{Average Number of Neighbors}} ($N_{N}$): computes the number of nodes belonging to the local area network of a specific node over time, which represents its neighborhood. $N_R$ corresponds to the average of the sum of all node neighbors at each time interval, $j = t_s$, when we compute this neighborhood, and the total number of time intervals, $t_s$, throughout all repetitions ($N_R$). We obtain $N_{N}$ through~(\ref{eq:nNeigh}).

\begin{equation}
\footnotesize
N_N = \mathlarger{\sum}\limits_{i\;=\;1}^{N_R} \; \mathlarger{\sum}\limits_{j\;=\;1}^{t_s} \; \dfrac{ N_{ij}}{t_s \; \times \; N_R}
\label{eq:nNeigh}
\end{equation}

\textbf{\textit{Average Number of Health Communities}} ($N_{C}$) accounts the average number of communities established by a node across time. $N_{C}$ corresponds to the average of the sum of all health communities formed by a node throughout all repetitions ($N_R$), as presented in~(\ref{eq:nC}).

\begin{equation}
\footnotesize
N_{C} = \mathlarger{\sum}\limits_{i\;=\;1}^{N_R} \; \mathlarger{\sum}\limits_{j\;=\;1}^{t_s} \; \dfrac{ C_{ij}}{t_s \; \times \; N_R}
\label{eq:nC}
\end{equation}

\textbf{\textit{Hit Rate}} ($HR$) indicates the rate of success for delivering data to the appropriate person. $HR$ corresponds to the ratio of the total hits to sensitive data ($A_{Sucess}$) to the total number of times sensitive data available to be accessed ($A_{Disp}$), and it's obtained through~(\ref{eq:HR}).

\begin{equation}
\footnotesize
HR = \dfrac{A_{Success}}{A_{Disp}} \times 100
\label{eq:HR}
\end{equation}

\textbf{\textit{Fault Rate}} ($FR$) accounts data disseminated and not accessed by other nodes. It's the percentage of data protected in an emergency and calculated by~(\ref{eq:faultRate}).

\begin{equation}
\footnotesize
FR = 100 - HR
\label{eq:faultRate}
\end{equation}

\textbf{\textit{Hit Rate by Competence}} ($HR_{Skill}$)
is equivalent to
$HR$, but it evaluates hit rate for each competence individually, according to competences seen in
Table~\ref{tab:aspectosAtribuidos}.
$HR_{Skill}$ is the
ratio between the hit rate of a single competence, $A_{Skill}$, and total hits to sensitive data ($A_{Sucess}$). It's obtained through~(\ref{eq:HRSkill}).

\begin{equation}
\footnotesize
HR_{Skill} = \dfrac{A_{Skill}}{A_{Success}} \times 100
\label{eq:HRSkill}
\end{equation}

\textbf{\textit{Average Time to Access Sensitive Data}} ($AT$) computes the time to access sensitive data in an emergency. $AT$ corresponds to the sum of the ratio from the time difference of ($t_r$) and the time of their dissemination ($t_d$), and the total of repetitions ($N_R$). It's calculated by~(\ref{eq:AT}).

\begin{equation}
\footnotesize
AT = \mathlarger{\sum}\limits_{i\;=\;1}^{N_R} \; \dfrac{t_{r(i)} - t_{d(i)}}{N_R}
\label{eq:AT}
\end{equation}

\subsection{Results and Analysis}\label{sec:results}

This subsection presents the results obtained during the simulations performed for each evaluated scenario. We analyze them in terms of STEALTH dependability (availability and reliability) and safety (availability). The STEALTH transient state ended at 25 sec of each repetition. This was the moment that the system has reached its
steady-state performance (i.e., when all nodes were able to send and receive messages). Hence, the transient removal has encompassed the deletion of results obtained until 25s of each repetition.

\subsubsection{Assessing the effect of node mobility}\label{sec:behavior}

\begin{figure*}[!t]
    \centering
    \subfloat{\includegraphics[width=120mm]{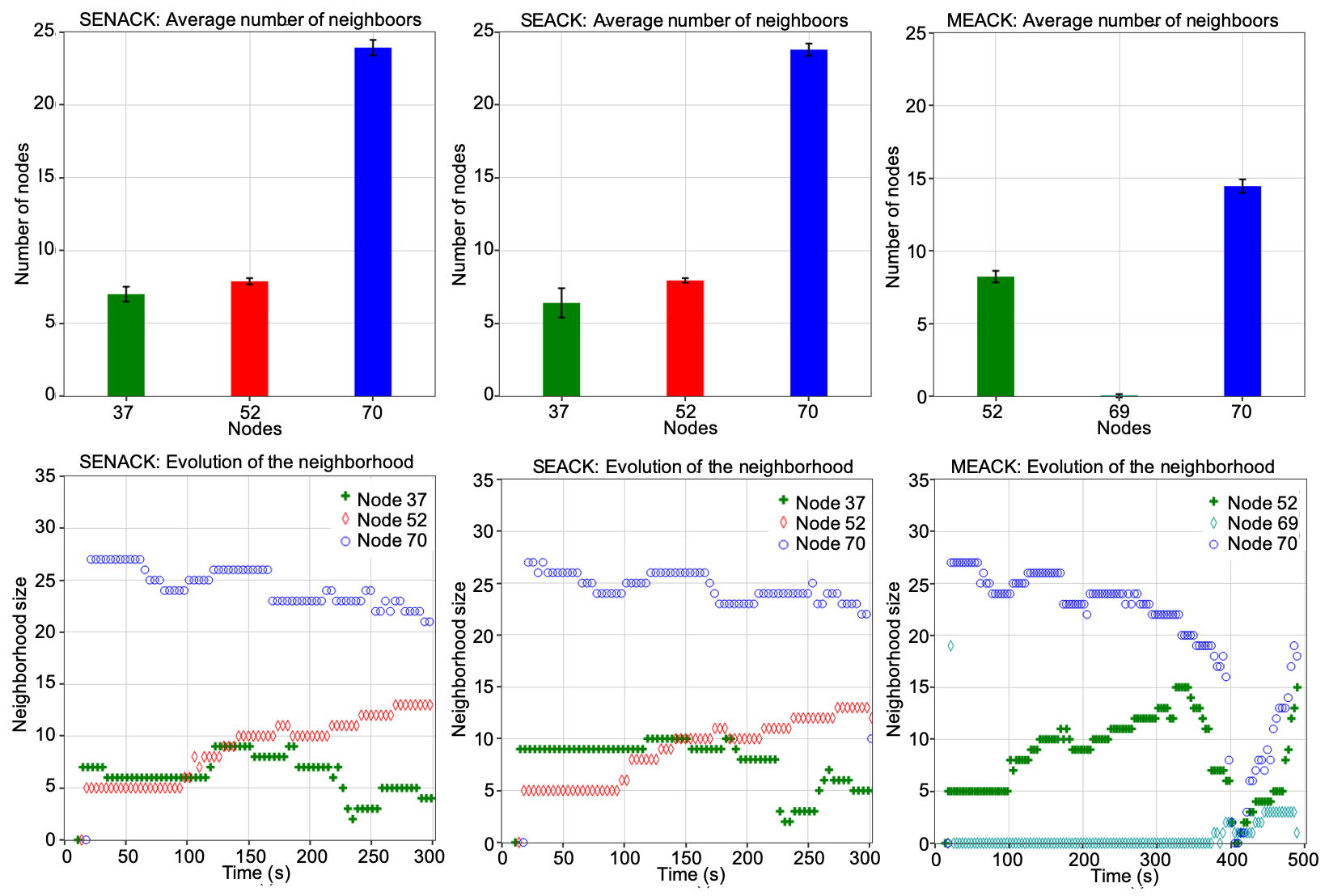}}
    \caption{Dynamics and size of local networks over time}
    \label{fig:dynamics}
    \vspace{-0.3cm}
\end{figure*}

The applied mobility model comprises pedestrians, and each node always takes the same path in all simulation repetitions~\cite{kth-walkers-20140505}. The average number of neighboring nodes ($N_{N}$) characterizes the model by the nodes previously selected in each scenario and Fig.~\ref {fig:dynamics} presents their neighborhood
behavior. We verify the neighborhood evolution along each repetition until a determined chosen time until the moment nodes enter in an emergency. Fig.~\ref {fig:dynamics}~top compares the behaviors of the neighborhoods of nodes 37, 52 and 70. While nodes 37 and 52 keep a similar $N_{N}$ in the SENACK and SEACK scenarios, node 70 neighborhood stands out for the number of nodes. MEACK scenario brings out different results, as demonstrates
Fig.~\ref {fig:dynamics}~top~right, where Node 69 presents a very small neighborhood ($N_N = 0 $).

The average number of neighbors ($N_N$) shows the mobility model and indicates the presence of a node neighborhood. But, it does not demonstrate neighborhood evolution across time. Fig.~\ref {fig:dynamics}~down
show the evolution of the neighborhood for nodes 37, 52, 69 and 70 in the evaluated scenario and in a given simulation repetition. The presence of the node 70 neighborhood in SENACK scenario, for example, indicates that STEALTH has created local networks around that node for 100\% of the time it was active in the simulation repetition (Fig.~\ref {fig:dynamics}~down~left). The opposite has happened with node 69 in scenario MEACK (Fig.~\ref {fig:dynamics}~down~right). This node starts creating local networks around it
at \mbox{393 sec} from the beginning of the simulation until \mbox{485 sec} (i.e., during 14.6\% of its running simulation time), when it entered in an emergency. In both cases, this behavior results from the employed mobility model.

The moment chosen to an emergency to happen influences the node neighborhood size. For instance, Fig.~\ref {fig:dynamics}~down~left shows the node 70 neighborhood composed of 22 nodes at \mbox{300 sec} of the SENACK scenario. Hence, node 70 could disseminate its sensitive data to several neighbors during an emergency. The MEACK scenario presents a distinct behavior. Fig.~\ref {fig:dynamics}~down~right shows the neighborhood evolution around nodes 52, 69 and 70. Node 69 established local networks only at time 393s from the simulation, over $14.6\%$ of its uptime until it stops the operation at time \mbox{485 sec}. Hence, it has maintained a tiny neighborhood ($N_{N} = 0$), and at that time, it had only 2 neighbors. In the majority of the simulation time, its mobility takes it away from the other nodes.

\subsubsection{Availability}
\label{sec:availability}

\begin{figure*}[!t]
    \centering
    \subfloat{\includegraphics[width=120mm]{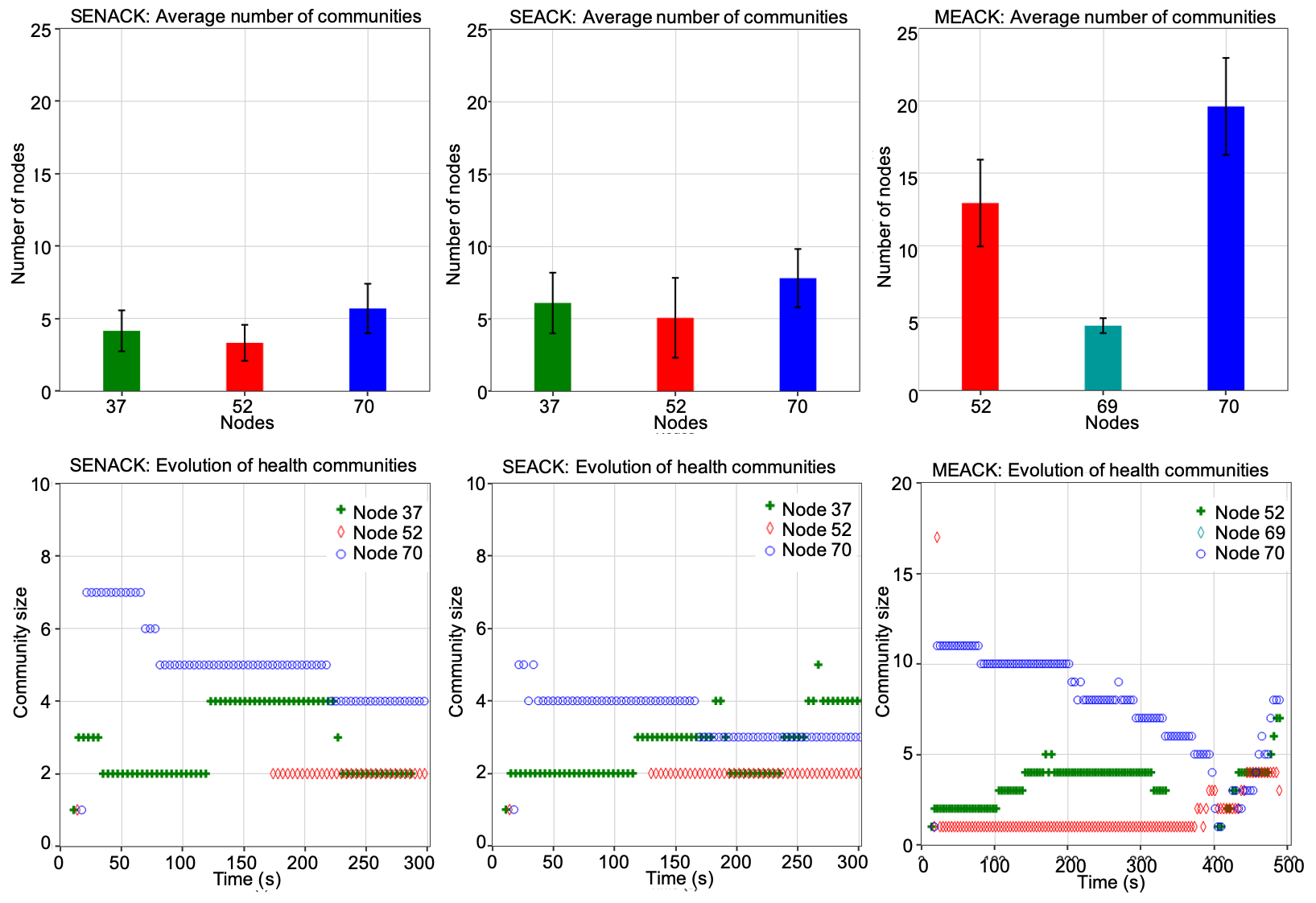}}
     \vspace{-0.3cm}
    \caption{Availability of health communities over time}
    \label{fig:availabitity}
    \vspace{-0.5cm}
\end{figure*}

We evaluate STEALTH availability (i.e., how successful STEALTH is in efficiently disseminating people sensitive data in an emergency.) Fig.~\ref{fig:availabitity}~top presents system behavior by synthesizing the average number of health communities ($N_{C}$) created over time in each scenario. Although SENACK and SEACK behave similarly, we observe a distinct behavior in each evaluated scenario. Node 37, for instance, establishes $N_C = 4$ in all simulations in SENACK. The best performance in all scenarios was achieved by node 70, but in MEACK it stands out for establishing almost 20 communities on average in each simulation. This behavior improves the system availability for the dissemination of sensitive data in emergencies. $N_C$ characterizes the dynamics of established local networks, mainly their topology. As expected, nodes mobility through distinct paths and their social aspects - \textit{interests} - significantly impacts on the creation of health communities. The dynamics of the evaluated health communities and their size over time in a specific simulation repetition is depicted in Fig.~\ref{fig:availabitity}~down. The mobility model made it possible interactions between nodes until the moment they entered in an emergency. Results show that STEALTH follows the dynamics of the established local networks, mainly owing to nodes mobility. \mbox{STEALTH} has verified changes in the node neighborhood and keep its health communities up to date. In the SENACK and SEACK scenarios, nodes 37, 52, and 70 have maintained health communities for 100\% of the time they were active. Over this time, STEALTH was always ready to disseminate node sensitive data, because there were other nodes assisting them. In the MEACK scenario, one observes a distinct behavior, as shown in Fig.~\ref{fig:availabitity}~right. Node 52 keeps health communities over 93.39\% of the time it was active. Upon entering an emergency, it has created a community with a couple of its neighbors, and successfully disseminated its sensitive data. Node 69 maintains a distinct behavior and keeps communities only over 11.30\% of the time until it enters an emergency, Finally, we observe that node 70 has maintained health communities the longest, 98.26\%. In all scenarios, STEALTH has identified node neighborhoods across the time. It successfully creates health communities and keeps them available to support a sensitive data dissemination in an emergency.

The size of communities is smaller than or equal to the size of the neighborhood at the same instant of time.
Health communities contemplate nodes around who are health-interested. Node 69 had been in a critical condition in MEACK scenario. It has created its first health community only at the time 393s, (Fig.~\ref{fig:availabitity} -- right). Until that moment, node 69 could not disseminate its sensitive data. 

\subsubsection{Reliability}\label{sec:reliability}

We analyze the STEALTH reliability in disseminating sensitive data from people in an emergency, We have conducted this analysis by evaluating the behavior of nodes 37, 52, 69, and 70 in each evaluated scenario. In the SENACK scenario, Table~\ref{tab:taxaSucesso} shows that node 70 successfully disseminates ($HR$) its sensitive data in 100\% of the emergencies throughout all repetitions. Nodes 37 and 52 have achieved a slighter difference in results than node 70, when they have disseminated their sensitive data in 97.14\% and 94.29\% of emergencies, respectively. In the SEACK scenario, node 70 was successful in 85.71\% of emergencies. Although all nodes in the MEACK scenario were 100\% successful in disseminating their data, this result was expected given the chosen emergency time. Grouping nodes in CoIs impacts on hit rate ($HR$), because clusters ensure the dissemination of sensitive data to specific nodes inside the community. We observe the importance of CoI in controlling data dissemination through the \textit{fault rate} ($FR$) metric. In the SEACK scenario, the sensitive data from node 37 was successfully accessed by other nodes in 25.71\% of emergencies. This behavior is due to the lack of a health community at these moments, or the disruption of its connection to other nodes due to their mobility.

\begin{table}[!t]
\caption{Data dissemination}
\vspace{0.1cm}
\label{tab:taxaSucesso}
\footnotesize
\setlength{\tabcolsep}{4pt}
\setlength{\extrarowheight}{2.0pt}
\begin{tabular}{l|l|cc|cc|cc}
\hlineB{2}
\multicolumn{2}{l|}{\textbf{Scenario}} & \multicolumn{2}{c|}{\textbf{SENACK}} & \multicolumn{2}{c|}{\textbf{SEACK}} & \multicolumn{2}{c}{\textbf{MEACK}} \\ \hline
\multicolumn{2}{l|}{\textbf{Metric}} &\textbf{$HR$} (\%)&\textbf{$FR$} (\%)&\textbf{$HR$}(\%)&\textbf{$FR$} (\%)&\textbf{$HR$} (\%)&\textbf{$FR$} (\%) \\ \hline
\multirow{4}{*}{\textbf{Node}} &37&97.14&2.86&74.29&\textcolor{blue}{\textbf{25.71}}&-&- \\
&52&94.29&5.71&77.14&22.86&\textcolor{blue}{\textbf{100}}&0 \\
&69&-&-&-&-&\textcolor{blue}{\textbf{100}}&0 \\
&70&\textcolor{blue}{\textbf{100}}&0&85.71&14.29&\textcolor{blue}{\textbf{100}}&0 \\ \hlineB{2}
\end{tabular}
\vspace{-0.2cm}
\end{table}

The cost to access the sensitive data, disseminated over time, is represented by the average time to access it (\textbf{$AT$}), which is influenced by the dynamics of the established local area networks. Table~\ref{tab:tempoMedio} shows that the results mostly meet the IEEE maximum latency for medical alert delivery - 125 ms~\cite{ieee2012}. In the SENACK and SEACK scenarios, nodes have promptly accessed sensitive data from node 37 ($AT <$ 1 ms). However, node 70 achieves a higher cost in SEACK scenario, when the access to its sensitive data happens after 170 ms of its dissemination. In all other situations, STEALTH obtains a cost lower than 27 ms. CoIs contribute to the decision making in verifying the appropriate node to receive the data and to a lower cost to access the disseminated sensitive data.

\begin{table}[!t]
\centering
\caption{Latency in access to disseminated data}
\vspace{0.1cm}
\label{tab:tempoMedio}
\footnotesize
\setlength{\tabcolsep}{10pt}
\setlength{\extrarowheight}{2.0pt}
\begin{tabular}{l|l|c|c|c}
\hlineB{2}
\multicolumn{2}{l|}{} & \multicolumn{3}{c}{\textit{Average Time to Access Sensitive Data} (ms)} \\ \hline
\multicolumn{2}{l|}{\textbf{Scenario}} &\textbf{SENACK}&\textbf{SEACK}&\textbf{MEACK}\\ \hline
\multirow{4}{*}{\textbf{Node}}  
& 37  & \textcolor{blue}{\textbf{$<$ 1}} & \textcolor{blue}{\textbf{$<$ 1}} & - \\
& 52  & 2.5 & 3 & \textcolor{blue}{\textbf{4}} \\
& 69  & - & - & \textcolor{blue}{\textbf{$<$ 1}} \\
& 70  & 17 & \textcolor{blue}{\textbf{170}} & 27 \\ \hlineB{2}
\end{tabular}
\vspace{-0.4cm}
\end{table}

Social aspects (i.e., \textit{interests} and \textit{competences}) in association with CoIs allow to assess the trust level of nodes and enable controlling their sensitive data dissemination. This dissemination occurs only to nodes belonging to a health community and in light of the node competence (Table~\ref{tab:aspectosAtribuidos}). This process succeeds under a \textit{zero-knowledge} condition (i.e., regardless node previous interactions). The relevance of the node competence is evaluated by the metric \textit{successful access to data by competence} ($HR_ {Skill}$), as shown in Table~\ref{tab:taxaMedia}. In the SEACK scenario, 76.92\% of the total sensitive data dissemination went to nodes with other competences. In 50\% of sensitive data dissemination from node 52, there was a node with doctor competence to access its data. This behavior indicates that in 50\% of emergencies, \mbox{STEALTH} found out the presence of at least one doctor in the available health community. The success observed in the MEACK scenario (100\% of data disseminated to nodes with doctor competence) is expected since both competences and emergency time contribute to this.

\begin{table}[!t]
\centering
\caption{Dissemination control}
\vspace{0.1cm}
\label{tab:taxaMedia}
\footnotesize
\setlength{\tabcolsep}{2pt}
\setlength{\extrarowheight}{2.0pt}
\begin{tabular}{l|l|ccc|ccc|ccc}
\hlineB{2}
\multicolumn{2}{l|}{\textbf{Scenario}} & \multicolumn{3}{c|}{\textbf{SENACK}} & \multicolumn{3}{c|}{\textbf{SEACK}} & \multicolumn{3}{c}{\textbf{MEACK}}\\ \hline
\multicolumn{2}{l|}{\textbf{Node}} &37&52&70&37&52&70&52&69&70\\ \hline
\multirow{4}{*}{\textbf{$HR_{Skill}$}}  & Doctor &44.12&81.82&\textcolor{blue}{\textbf{91.43}}&7.69&\textcolor{blue}{\textbf{50}}&18.18&\textcolor{blue}{\textbf{100}}&\textcolor{blue}{\textbf{100}}&\textcolor{blue}{\textbf{100}}\\
                   & Nurse  & 32.35&18.18&8.57&7.69&30&13.63&0&0&0 \\
                   & Caregiver  &23.53&0&0&7.69&16.67&36.36&0&0&0 \\
                   & Other  &0&0&0&\textcolor{blue}{\textbf{76.92}}&3.33&31.81&0&0&0\\ \hlineB{2}
\end{tabular}
\vspace{-0.55cm}
\end{table}

\begin{figure*}[!t]
    \centering 
    \subfloat{\raisebox{7mm}{\includegraphics[width=30mm]{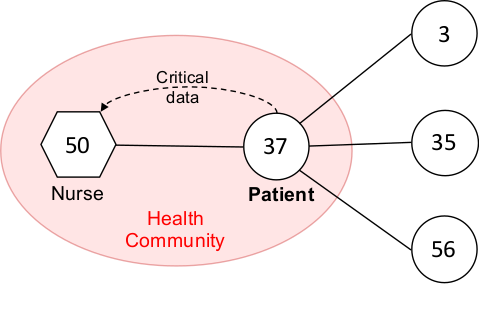}}\label{fig:snap_37_v1_300s}}
    ~
    \subfloat{\includegraphics[width=36mm]{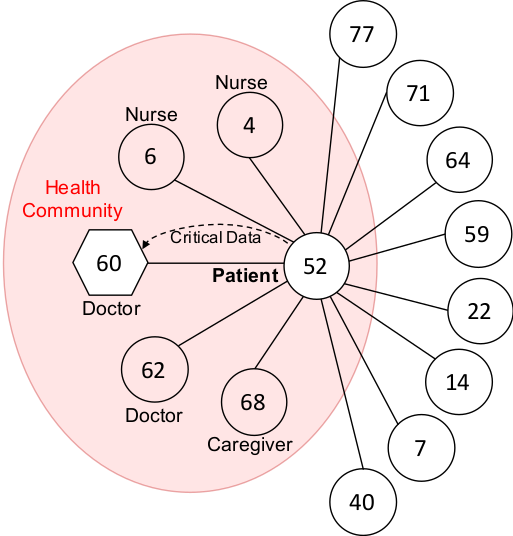}}\label{fig:snap_52_v1_300s}
    ~
    \subfloat{\includegraphics[width=42mm]{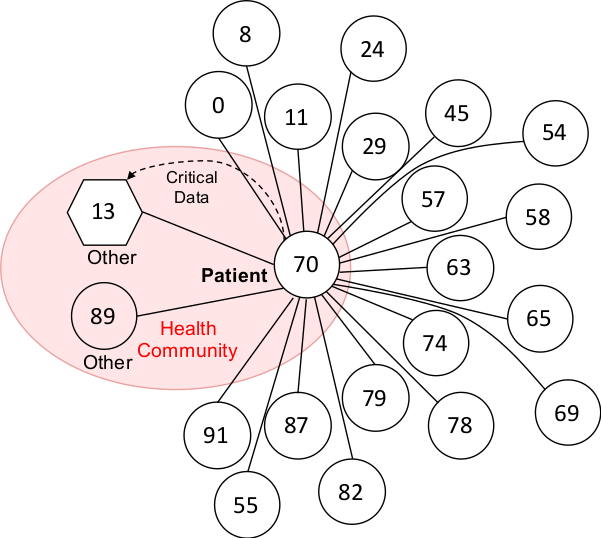}}\label{fig:snap_70_v1_300s}
    \vspace{-0.4cm}
    \caption{SENACK scenario: Nodes 37, 52 and 70 health communities on a critical event at \mbox{300 sec} of Simulation}
    \label{fig:snap_v1_300s}
    \vspace{-0.5cm}
\end{figure*}

The network topology of the SENACK scenario in one specific simulation is depicted in Fig.~\ref{fig:snap_v1_300s}. We observe the neighborhood of the evaluated nodes and their health community. There was only one neighbor inside node 37 health community (i.e., node 50) who held the nurse competence (Fig.~\ref{fig:snap_v1_300s}~left). As nodes 3, 35, and 56 were not health-interested, node 37 has disseminated its sensitive data to node 50. The node 52 neighborhood was more significant and impacted in its health community size. This condition has increased the likelihood of node 52 health community having members with different skills, as shown in Fig.~\ref{fig:snap_v1_300s}~middle. There are two nodes with doctor competence (i.e., nodes 60 and 62), the first one was the appropriate to access sensitive data from node 52. Although both possess the same competence, their interests were distinct. The number of interests in common between nodes 60 and 52 is greater than between 60 and 62. Hence, node 60 achieves a higher trust level and it is selected. Something similar happens to node 70, which its neighborhood is the biggest (Fig.~\ref{fig:snap_v1_300s}~right). However, its health community comprises only two neighboring nodes (i.e., nodes 13 and 89), and they both possess the \textit{other} competence. Hence, the selection of the node to disseminate the sensitive data occurs based on the number of common interests they had each other. In this case, node 13 is selected. Sensitive data is disseminated in a controlled manner, with no exposure of the data to unauthorized people. The decision-making process for choosing nodes to disseminate data is similar in the other scenarios.

\vspace{-0.3cm}
\section{Conclusion and future work}\label{sec:conclusion}

This article presented \mbox{STEALTH}, a system for disseminating sensitive health data in a controlled manner in dynamic wireless local area networks. It builds virtual clusters taking into account communities of interest and it employs social trust to enable the devices to decide on data dissemination under an emergency. Simulation results from realistic scenario have shown STEALTH ability to ensure the dissemination of sensitive data. STEALTH has achieved a reliability of up to 97.14\% in the access to disseminated data, a maximum latency of \mbox{170 ms}, and up to 100\% of availability. As future work, the reliability in decision-making will be contrasted under security threats and simultaneous severe events to complement the safety vision presented in this work. 

\vspace{-0.3cm}
\bibliographystyle{IEEEtran}
\bibliography{IEEEabrv,sample.bib}

\vspace{-1.5cm}
\begin{IEEEbiography}[{\includegraphics[width=1in,height=1.25in,clip,keepaspectratio]{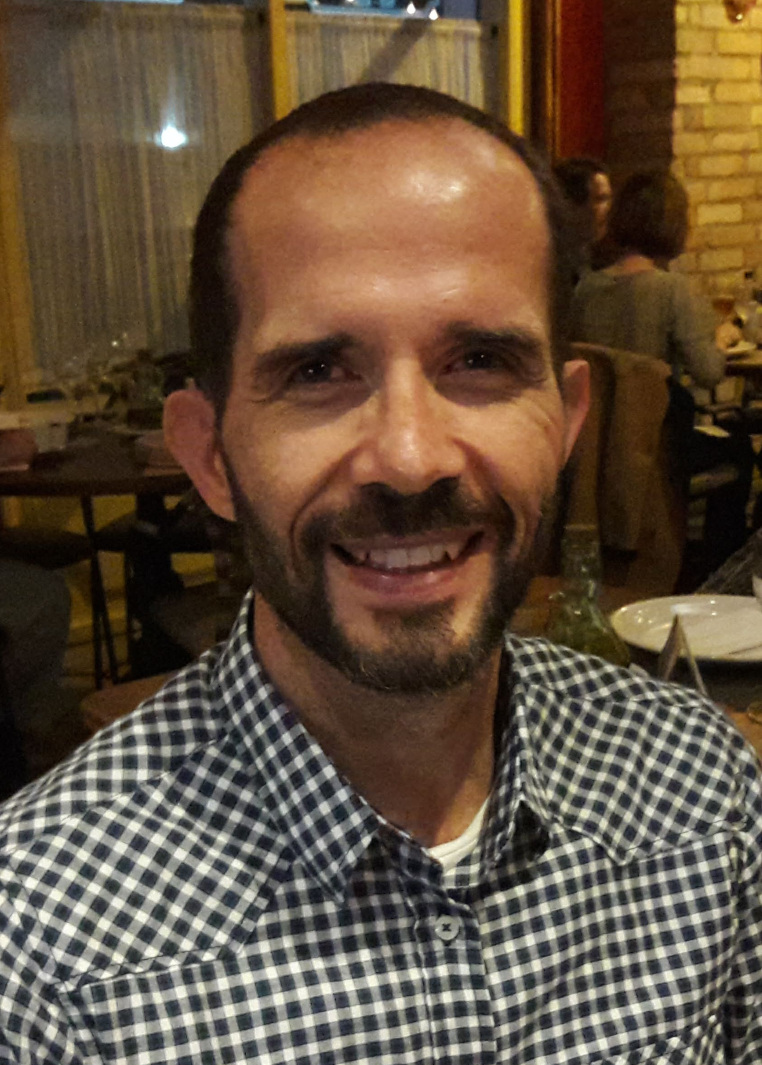}}]{Agnaldo de Souza Batista} received the B.E. degree from Catholic University of Pelotas, Brazil, and is Master M.S. in Informatics from Federal University of Parana (UFPR). He is now pursuing his Ph.D. degree in computer science at UFPR. His research interests are robust systems, data security, wireless networking, Internet of Things (IoT), \mbox{e-health}, and management of critical events, Agnaldo is member of Brazilian Computer Society (SBC).
\vspace*{-5\baselineskip}
\end{IEEEbiography}

\begin{IEEEbiography}[{\includegraphics[width=1in,height=1.25in,clip,keepaspectratio]{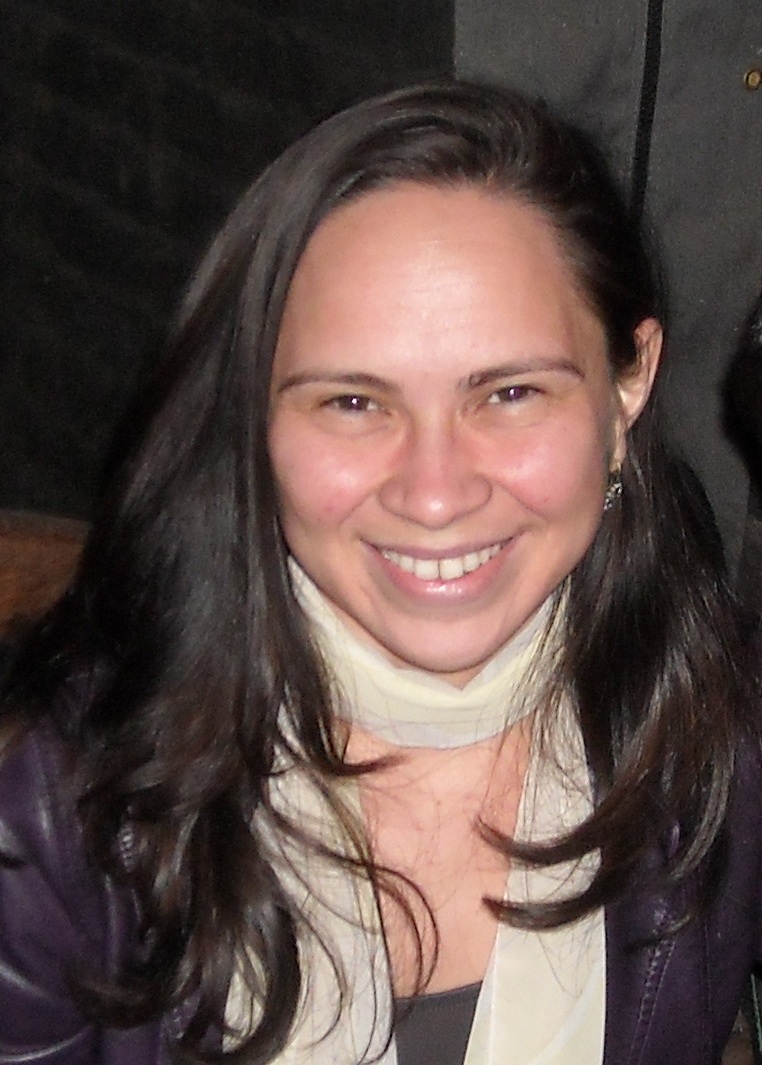}}] {Michele Nogueira} is professor of computer science at Federal University of Parana, where she has been since 2010. She received her doctorate in computer science from the University Pierre et Marie Curie – Sorbonne Universites, Laboratoire d’Informatique de Paris VI (LIP6) in 2009. She was a Visiting Researcher at Georgia Institute Technology (GeorgiaTech) and a Visiting Professor at University Paul Sabatier in 2009 and 2013, respectively. Her research interests include wireless networks, security and dependability. She has been a recipient of Academic Scholarships from Brazilian Government on her undergraduate and graduate years, and of international grants such as from the ACM SIGCOMM Geodiversity program. She is also Associate Technical Editor for the IEEE Communications Magazine and the Journal of Network and Systems Management.
\vspace*{-5\baselineskip}
\end{IEEEbiography}

\begin{IEEEbiography}[{\includegraphics[width=1in,height=1.25in,clip,keepaspectratio]{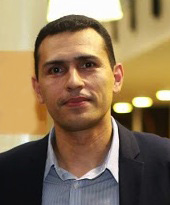}}]{Aldri L. dos Santos} is professor of the Department of Informatics at Federal University of Paraná (UFPR). Aldri is PhD in Computer Science from the Federal University of Minas Gerais, Master in Informatics and Bachelor of Computer Science at UFPR. Aldri is member of the faculty of the Graduate Program in Informatics (PPGInf) UFPR, working in the following research areas: network management, fault tolerance, security, data dissemination, wireless ad hoc networks and sensor networks. He is leader of the research group (Wireless and Advanced Networks). Aldri has also acted as reviewer for publications as IEEE ComMag, IEEE ComNet, ComCom, IEEE Communications Surveys and Tutorials, IEEE eTNSM, JNSM, Ad hoc Networks. Aldri has served as member of the technical committee of security information and IEEE Communication Society Communication (ComSoc).
\end{IEEEbiography}

\end{document}